\documentclass[preprint2]{aastex}

\slugcomment{To appear in ApJ}

\shorttitle{Gamma-Ray Emission from 3C\,111}
\shortauthors{Hartman et al.}

\begin{document}

\title{Gamma-Ray Emission from the Broad-Line Radio Galaxy 3C\,111}

\author{R. C. Hartman\altaffilmark{1},
        M. Kadler\altaffilmark{2,3,4},
	and
	J. Tueller\altaffilmark{1}
       }
\altaffiltext{1}{Astrophysics Science Division, NASA's Goddard Space Flight 
Center, Greenbelt, MD 20771, USA}
\altaffiltext{2}{Dr. Karl Remeis-Observatory, University of Erlangen-Nuremberg, 
Sternwartstrasse 7, 96049 Bamberg, Germany}
\altaffiltext{3}{CRESST/NASA Goddard Space Flight Center, Greenbelt, MD 20771, USA}
\altaffiltext{4}{Universities Space Research Association, 10211 Wincopin Circle, Suite 500, Columbia, MD 21044, USA}

\email{Robert.C.Hartman@nasa.gov, matthias.kadler@sternwarte.uni-erlangen.de, 
Jack.Tueller@nasa.gov}

\begin{abstract}
The broad-line radio galaxy 3C\,111 has been suggested as the counterpart 
of the $\-$--ray source 3EG\,J0416+3650.  While 3C\,111 meets most of the 
criteria for a high-probability identification, like a bright flat-spectrum 
radio core and a blazar-like broadband SED, in the Third EGRET Catalog, the 
large positional offset of about 1.5$^\circ$ put 3C\,111 outside the 99\% 
probability region for 3EG\,J0416+3650, making this association 
questionable.  We present a re-analysis of all available data for 
3C\,111 from the EGRET archives, resulting in probable detection of 
high-energy $\gamma$--ray emission above 1000\,MeV from a position 
close to the nominal position of 3C\,111, in two separate viewing periods 
(VPs), at a 3$\sigma$ level in each.  A new source, GRO\,J0426+3747, appears to
be present nearby, seen only in the $>$1000\,MeV data.  For $>$100\,MeV, the 
data are in agreement with only one source (at the original catalog position)
accounting for most of the EGRET-detected emission of 3EG\,J0416+3650.  A 
follow-up \textit{Swift} UVOT/XRT observation reveals one moderately bright 
X--ray source in the error box of 3EG\,J0416+3650, but because of the large 
EGRET position uncertainty, it is 
not certain that the X--ray and $\gamma$--ray sources are associated.  A
\textit{Swift} observation of GRO\,J0426+3747 detected no X--ray source nearby.
\end{abstract}


\section{Introduction}
One of the main scientific goals of the recently-launched $\gamma$--ray 
astronomy satellite mission GLAST and its Large Area Telescope (LAT) 
\citep{Ri07, Mi07} is to shed light on the 
nature of powerful relativistic extragalactic jets, which are 
ejected from the nuclei of some active galaxies (AGN).  Originally, this class 
of AGN was defined based on the bright and prominent radio emission from these
jets: the radio-loud population of active galaxies.  Based on data from EGRET, 
the high-energy $\gamma$--ray telescope on the {\it Compton Gamma Ray 
Observatory}, it was realized \citep{Fi94, Th94, vM95} that the largest 
population of extragalactic $\gamma$--ray sources in the GeV regime is 
represented by those radio-loud AGN whose jets are pointed at a small angle 
to the line of sight.  An intimate link is tied between radio VLBI and 
high-energy $\gamma$--ray astronomy by the fact that the bright compact radio 
emission of these so-called blazars provides excellent 
targets for parsec-scale resolution VLBI observations of their jet structure.

Of special interest are the questions of where in the AGN jets the bright 
$\gamma$--ray emission is produced, and how the emission is interacting with its 
immediate environment and with other parts of the jet.  This knowledge would 
enable us to put crucial constraints on the processes of jet formation, 
collimation, and acceleration.  GLAST is expected to yield densely sampled
$\gamma$--ray light curves of hundreds of extragalactic jets that are bright 
enough to be detected on time scales of days to weeks, and thousands on time 
scales of months to years.  Most of these objects will be blazars; in fact, all 
but two of the firmly identified extragalactic EGRET sources (LMC, Sreekumar et 
al. 1992; Centaurus\,A, Sreekumar et al. 1999) are blazars.  In contrast to 
the blazars, most radio galaxies have larger inclination angles.  That allows 
better (deprojected) linear resolution with VLBI observations, and in the case 
of stratified jet structures, it allows observations of the slower jet layers, 
e.g. a sheath, whose emission may be swamped by the much brighter beamed 
emission from faster jet regions, e.g. a fast spine, in blazars.  \citet{Ghi05} 
have presented such a spine-sheath stratified-jet model, predicting detectable 
$\gamma$--ray emission from the nuclei of some radio galaxies.  In their model, 
each of the two emission regions sees photons coming from the other part 
relativistically enhanced because of the relative speed difference, giving rise 
to an additional inverse-Compton emission component.

In this letter we provide additional support for the identification of the 
broad-line radio galaxy 3C\,111 as a $\gamma$--ray source, responsible for a 
portion of the source 3EG\,J0416+3650.  Such an association was suggested as 
possible in the third EGRET catalog \citep[3EG]{Ha99}, but was considered 
unlikely because of the 
large positional offset of 3C\,111 from 3EG\,J0416+3650.  The present work was 
stimulated by the recent report of \citet{Sg05}, who used multiwavelength data 
to strengthen the case for the association between 3C\,111 and 3EG\,J0416+3650.

We show here that 
3EG\,J0416+3650 is most likely composed of at least two, and more likely three, 
separate sources, one of which is in good positional agreement with 3C\,111.  
As demonstrated recently, based on VLBA monitoring data at 
$\lambda 2$\,cm from the MOJAVE program \citep{Kad08}, the parsec-scale 
jet of 3C\,111 shows a variety of physically different regions in a 
relativistic extragalactic jet, such as a compact core, superluminal jet 
components, recollimation shocks, and regions of interaction between the jet and 
its surrounding medium, which are all possible sites of $\gamma$--ray production.  
Its relatively large inclination angle of $\sim$19$^\circ$ makes 3C\,111 a 
particularly well-suited target for tests of structured-jet models, such as 
the model of \citet{Ghi05}.  We describe our re-analysis of the available EGRET 
data on 3C\,111, as well as follow-up \textit{Swift} OVOT and XRT observation, 
in Sect.~\ref{sect:analysis}, and discuss our
results and their implications in Sect.~\ref{sect:conclusions}.

\section{Analysis}
\label{sect:analysis}
\subsection{EGRET Data Analysis}
Using detailed analysis of reprocessed EGRET data\footnote{The EGRET data 
reprocessing consisted of: 1) Manual examination of all events with energy 
$>$1000\,MeV which had previously been rejected, adding a considerable number of 
events to the database; 2) adjustment of the EGRET sensitivity tables 
\citep{Ber01}; 3) complete re-analysis of one viewing period (not used here) 
because of numerous errors in the original event selection.},
we have investigated the possibility that 3EG\,J0416+3650 is a superposition 
of at least two sources, one being at or near the position of 3C\,111.  The 
analysis for the 3EG catalog was based on photons with energies $>$100\,MeV, 
for which the 68\% containment angle of the spectrum-dependent point spread 
function (PSF) is always several degrees.  For energies $>$1000\,MeV, the 
half-angle of the cone containing 68\,\% of the photons from a source is 
considerably smaller, slightly less than one degree.  On the other hand, 
the number of photons with energies above 1000\,MeV 
is much smaller than that above 100\,MeV.  Both of those integral energy ranges 
were investigated for this study.

For each viewing period (VP) in which 3C\,111 and 3EG\,J0416+3650 were in the 
EGRET field of view (FoV), the following procedure was initially used:

\begin{enumerate}
\item Using the standard EGRET likelihood software ({\sc like}), two sources were 
modeled, one at the 3EG position for 3EG\,J0416+3650 and one at the nominal 
position of 3C\,111, and {\sc like} then provided the best estimates of the 
significance and flux of the two assumed sources.
\item If either of the two assumed sources had a test statistic (TS) of at least 
4.0 (roughly 2$\sigma$), {\sc like} was then allowed to optimize the 
significance of the two sources by moving their positions.
\end{enumerate}

The results from this procedure are shown in Table~\ref{tab:decomposition}, 
for the two energy ranges and the two assumed sources.  The first five columns 
show the VPs, the starting dates of the observations, the lengths of the VPs, 
and the off-axis angles of 3C\,111.  
The line showing VP=P1234 is the analysis of the summed data for the VPs included 
in 3EG, VP0002 through VP4270.  For each object and energy range, the table 
shows the maximum TS near the starting position, the change in 
position (from the starting point) to obtain the maximum TS, and the resulting 
photon flux in units of $10^{-8}$ photons\,cm$^{-2}$\,s$^{-1}$.  
Blank entries for TS indicate values of 0.  For 
TS values less than 4.0, there are no entries 
in the next two columns.  In addition to the VPs listed, VPs 0005, 0010, 0365, 
2130, 2210, 4120, and 4260 were tested, and showed no indication (TS=0) of 
emission from either 3C\,111 or 3EG\,J0416+3650; VP0390 also showed TS=0, but 
is shown in the table because it was the best exposure to the relevant sky 
region.  For the cases where the offsets are small compared with the PSF, there 
is little difference in the TS and flux at the maximum TS positions compared 
with those at the original positions.

Considering that seven VPs are not shown because they yielded TS=0 for all four 
source/energy cases, the density of detections in this table is very low, 
even with the low requirement of TS$\ge$4.0.  One of the 3C\,111 detections 
(VP0150 for E$>$1000\,MeV) has an offset position value larger than would be 
expected, and out of line with the other detections.  It is discussed below.

Two VPs (0002 and 3211) provide reasonably strong detection of 3EG\,J0416+3650,
which agrees with the information provided in 3EG.  The summed exposure P1234 
provides nearly the same significance of detection (but a much lower average 
flux than in VPs 0002 and 3211), indicating that the object was emitting 
$\gamma$ rays at a low level 
even when it was not significantly detectable in individual VPs.

There are five results that support 
the proposed 3C\,111 $\gamma$--ray source, two in E$>$100\,MeV and three in 
E$>$1000\,MeV.  With the exception of VP0150 at $>$1000\,MeV,  
the values of the position offsets are compatible 
with the position of 3C\,111.  Only one VP (0310) shows detections in both 
energy ranges.  Like 3EG\,J0416+3650, 3C\,111 appears to radiate $\gamma$ 
rays with a low duty cycle, occasionally becoming fairly bright.

The number of photons $>$100\,MeV from 3C\,111 was 64, 44 of them in VP0310 
and 14 in VP3215.  Only about 20 photons $>$1000\,MeV were detected from 
3C\,111, six in VP0150, four in VP0310, and four in VP3250.

For 3EG\,J0416+3650, there was a total of 152 photons $>$100\,MeV, 
37 of them in VP3211 (the most significant single detection) and 33 in VP0002.  
P1234 yielded seven photons above 1000\,MeV from 3EG\,J0416+3650, but as 
indicated in Table 1, they produced a TS of only 3.1, about 1.8$\sigma$.

We have further investigated a small region around 3EG\,J0416+3650 by generating 
$5^\circ \times 5^\circ$ likelihood maps, for the energy ranges $>$100\,MeV and 
$>$1000\,MeV, for each VP in which 3C\,111 and 3EG\,J0416+3650 were in the FoV.   
A sample of the more informative likelihood maps is shown in Figure 1.  Maps 
for the summed data, P1234, are shown in Figure 2.  The contours in the upper 
($>$100\,MeV) half of Figure 2 are similar to those in the corresponding figure 
from 3EG, but in the current version, the extension toward the 3C\,111 position 
is more pronounced.  We draw the following conclusions from these maps:

\begin{enumerate}
\item There appear to be at least three sources (possibly more) contributing, two 
in the $>$1000\,MeV energy range and at least one in the $>$100\,MeV range.  One 
of the two $>$1000\,MeV sources is very near the position of 3C\,111; for the 
other, which is of comparable strength, we find no obvious identification 
(see below).  In the following, we designate this source GRO\,J0426+3747; as 
seen from Figure 2, the 68\,\% confidence error on its position is about 0.8$^\circ$.  
\item Little or no significant emission is detected from the two $>$1000\,MeV 
sources in the lower energy band.
\item The $>$100\,MeV emission is dominated by a source near the 3EG catalog 
position. There is no significant detection in the $>$1000\,MeV energy range.  
In the following, the name 3EG\,J0416+3650 refers implicitly to this source.
\item The $>$1000\,MeV map for VP0150 shows why the position offset for 3C\,111 
is so large (0.43$^\circ$) in Table 1 in that VP.  There are only six photons 
that can be clearly attributed to 3C\,111 in that map.  One additional photon 
is nearly a degree away from 3C\,111.  This shifts the maximum likelihood 
position toward that single photon, which could be either from the tail of the 
PSF, or from the diffuse Galactic foreground.
\end{enumerate}

\begin{figure}[t!]
\centering
\hspace{-1.10cm}
\includegraphics[clip,width=0.85\columnwidth]{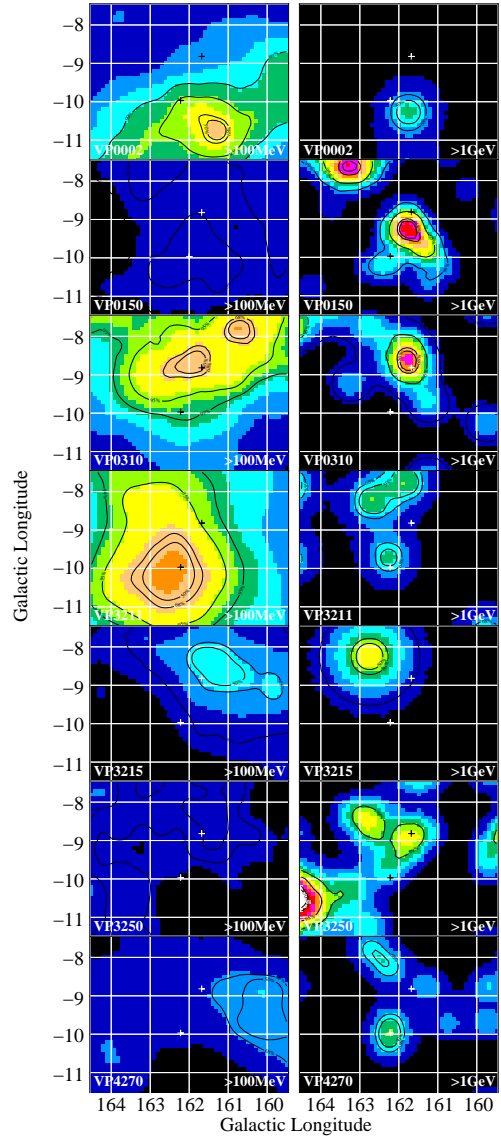}
\caption{Likelihood maps for the seven viewing periods with TS$>$3 (anywhere in
the map, either energy range).  The left column panels are for E$>$100\,MeV, 
while the right column panels show E$>$1000\,MeV, with fewer photons but a 
considerably smaller point spread function.  Crosses show the positions of 3C\,111 
and 3EG\,J0416+3650; ref. Figure 2.}
\end{figure}

\begin{figure}[t!]
\centering
\hspace{-1.1cm}
\includegraphics[clip,width=0.85\columnwidth]{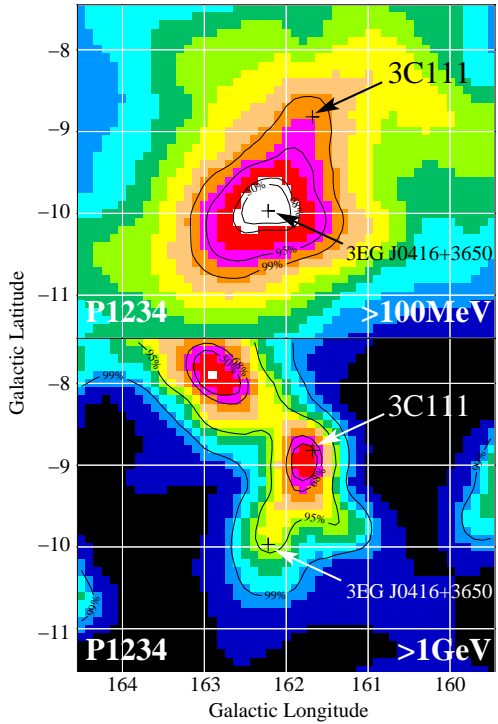}
\caption{Likelihood maps for the summed exposure of all viewing periods, for 
$>$100\,MeV (top) and $>$1000\,MeV (bottom).  On the scale used here, the 
position of Swift\,J041554.3+364926 is almost indistinguishable from that of 
3EG\,J0416+3650, and is therefore not shown.}
\end{figure}

As an additional check on the reality of the two $>$1000\,MeV sources, we have 
compared the number of occurrences of $\ge$2$\sigma$ excesses in 17 maps (5$^\circ$ 
by 5$^\circ$) with the statistically expected value.  The results are shown in 
Table 2.  For the regions of the maps away from the two ``source'' positions, the 
number of $>$2$\sigma$ excesses is about as expected from statistical fluctuations, 
taking into account the number of PSFs contained within each map.  Near the 
locations of the two $>$1000\,MeV sources, however, there are far more excesses 
than expected, which provides added confidence in the detections of 3C\,111 and 
GRO\,J0426+3747.  For each of VPs 0005 and 3215 (which had very low exposures), 
there was exactly one photon detected in the 5$^\circ$ x 5$^\circ$ region for 
which the TS maps were generated.  Both of those photons were very close to the 
GRO\,J0426+3747 position obtained from the summed emission in P1234, well within 
the 68\,\% PSF extent.  An examination of a somewhat larger region indicated that 
the probability of a photon from the diffuse emission that close to a particular 
point was less than 0.01.  Those two events are indicated with ``s'' in Table 2.

\subsection{\textit{Swift} Follow-Up X-Ray Observations and Their Analysis}
There are no known high-energy sources close to the nominal best-fit positions 
of 3EG\,J0416+3650 and GRO\,J0426+3747. Therefore, we have conducted follow-up 
optical/UV and X--ray observations with the UVOT and XRT instruments on the 
NASA satellite \textit{Swift} \citep{Geh04}.  The 3EG\,J0416+3650 field was 
observed for 10.4\,ksec on December 14, 2007\footnote{Observation ID 
00031057001}, and the GRO\,J0426+3747 field for 9.0\,ksec on 20 December, 
2007\footnote{Observation ID 00031072001}.  No significant X--ray source was 
detected above the background level in the 23.6$\times$23.6\,arcsec XRT FoV 
around GRO\,J0416426+3747.  The XRT observation of the 3EG\,J0416+3650 field
detected one $4.8 \sigma$ source at the position 
{\mbox 04\,h 15\,m 54.3\,s, +36\,d 49\,m 26\,s}, with an uncertainty of 5.8 
arcsec.

We extracted spectra from the source (which we designate 
Swift\,J041554.3+364926) and background regions using {\sc xselect}, and response 
and ancillary files provided by the \textit{Swift} calibration data base 
(CALDB)\footnote{\tt http://heasarc.gsfc.nasa.gov/docs/caldb/swift/}.  The low 
count rate of $0.0026\pm 0.0006$\,cts\,+s$^{-1}$ limits the ability to derive 
spectral information.  However, it is striking that only 4 photons (out of 26) 
were detected between 0.2\,keV and 1\,keV.  To characterize the spectrum, 
we used Cash statistics (Cash 1979) in {\sc xspec} and determined 
the maximum allowable power-law photon index in a power-law fit under the 
assumption of zero absorption, as may be expected from a very local source.  
The  result is slightly dependent on the data binning chosen, but suggests a 
very hard spectrum, with the maximum allowed photon index in the range 
1.1-1.3.  If the absorption parameter is allowed to vary freely during the 
fit, the photon index is poorly constrained, and steeper values up to 2.7 are 
statistically allowed. 
 
\subsection{Spectral energy distributions}
The historical spectral energy distribution (SED) of 3C\,111 is shown 
in Figure 3.  The millimeter flux range 
was determined with permission from the SMA Submillimeter Calibrator List 
archived data\footnote{\tt http://sma1.sma.hawaii.edu/callist/callist.html} 
\citep{Gur07}.  The R-band flux range has been determined from data obtained in 
a long-term monitoring program with the Liverpool telescope (PI: I. McHardy) 
between November 2004 and July 2007 (S. Jorstad, priv. comm.).

\begin{figure}[t!]
\hspace{-0.45cm}
\includegraphics[clip,width=1.15\columnwidth]{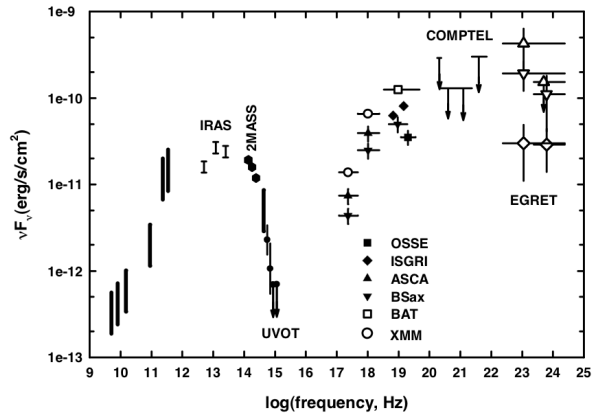}
\caption{Spectral energy distribution for 3C\,111.  The data points are 
non-contemporaneous; their values and references are tabulated in Table 3.  
Thick vertical lines indicate historical ranges for radio and submillimeter 
observations.  The thick vertical line in the optical indicates the range of 
R-band fluxes from 2004-2007 observations with the Liverpool Telescope - see 
text.  The NIR through UV points have been dereddened; see text.  For the EGRET 
data, the average values are indicated with open diamonds, the most 
significant single viewing period, VP0310, uses downward-pointing open triangles, 
and the maximum flux, VP3215, is shown with upward-pointing triangles.  The 
lowest upper limits for a single VP (not shown) are similar to the average.}
\end{figure}

The 3C\,111 SED is similar to those of EGRET-detected flat-spectrum radio 
quasars (FSRQs), except that the rollover in the NIR, optical, and UV is 
steeper here.  This makes the ``valley'' between the synchrotron and Compton 
``bumps'' appear deeper.  This could be due to absorption intrinsic 
to 3C\,111, since the data points have been corrected for Galactic absorption.

Figure 4 shows (a) a UVOT image of the Swift\,J041554.3+364926 region, with a 
circle superimposed indicating the 1--$\sigma$ position uncertainty for the 
X--ray source (as determined with the {\sc ftools} task {\sc xrtcentroid}); 
and (b) an SED containing the very limited multiwavelength data available
for this source.  The EGRET points 
shown are for comparison only; despite its proximity to the 3EG position for 
the $\gamma$--ray source, it is not certain that Swift\,J041554.3+364926 is 
actually the same object as 3EG\,J0416+3650, because the EGRET position 
uncertainty is quite large (about 0.63$^\circ$ in radius).

\begin{figure}[t!]
\centering
\hspace{-0.5cm}
\includegraphics[clip,width=0.91\columnwidth]{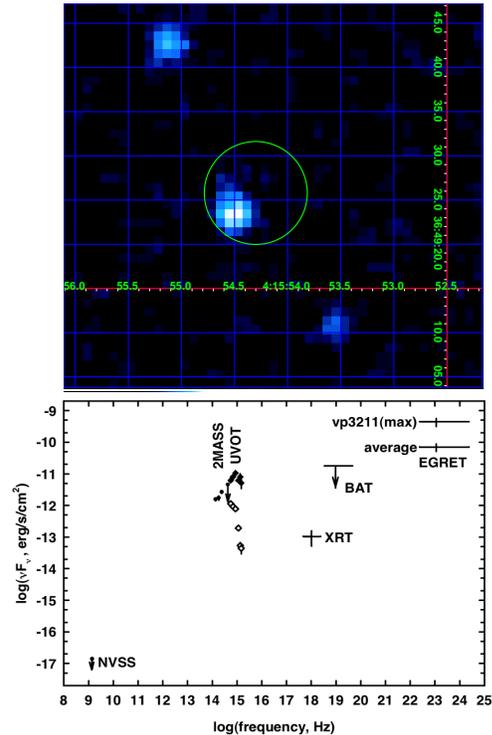}
\caption{(a) UVOT image (UW1 filter) of the region around Swift\,J041554.3+364926.  
The circle indicates the position uncertainty (68\% confidence) for the X--ray 
source; (b) Spectral energy distribution for Swift\,J041554.3+364926.  The data 
points are non-contemporaneous; their values and references are tabulated in 
Table 3.  For the UVOT data points, the observed values are 
shown as open circles, while dereddened values are shown as filled circles.  
The EGRET points may not be associated with the lower-frequency source - see text.}
\end{figure}

In the Swift\,J041554.3+364926 SED, the UVOT data are plotted twice, with and 
without dereddening, because it is not certain whether the object is nearby or 
extragalactic.  The open circles denote the observed fluxes; the dereddened 
fluxes are shown as filled circles.  Note that the fall-off in the optical-UV is 
extremely steep, especially in the observed fluxes.  If Swift\,J041554.3+364926 
is an extragalactic source, this can be attributed to absorption at the edge 
of a  foreground molecular cloud, part of the Taurus complex, with absorbing 
column density $N_{\rm H}=2.18 \times 10^{21}$\,cm$^{-2}$ \citep{Kal05}.  
NED shows $A_V$ of 1.893\,mag for the catalog position of the EGRET source, about 
0.1$^\circ$ away.  In the extragalactic scenario, Swift\,J041554.3+364926 must 
be a radio quiet source because of the very low upper limit on its radio flux 
density, 0.7 mJy at 1.4\,GHz, from the NVSS survey \citep{Con98}.  The shape 
of the dereddened UV/optical/IR range suggests an AGN ``blue bump'', usually 
attributed to thermal emission from an accretion disk.  However, without far IR 
and/or submillimeter data points, it is not possible to be certain of this 
interpretation.  Based on previous experience, a radio-quiet AGN is an unlikely 
identification for an EGRET source.

Alternatively, if Swift\,J041554.3+364926 is within our Galaxy, it would need to 
be an M star or a cataclysmic variable (CV) in order to produce the observed 
X--rays (R. Osten and S. Drake, priv. comm.).  In order to be 
visible optically, an M star would necessarily be in front of the molecular 
cloud, which is 140\,pc distant \citep{Eli78}, so the observed UVOT data points 
should be considered, rather than the dereddened ones.  The observed X--ray 
emission is stronger than expected from an M star except during an exceptional 
flare (R. Osten and S. Drake, priv. comm.).  If the object is a CV, 
its luminosity is much greater, so it must be behind the Taurus absorber, making 
the dereddened UVOT points appropriate; in that case the X--ray emission is 
reasonable.  Based on previous experience, neither an M star or a CV would be a 
likely identification for an EGRET source.

Table~\ref{tab:3c111_sed} and Table~\ref{tab:J0416_sed} provide the values 
plotted in Figures 3 and 4, with references.

\section{Discussion and Conclusions}
\label{sect:conclusions}
The $\gamma$--ray source 3EG\,J0416+3650 seems to be composed of at least three 
variable sources.  One of the sources, detected only above 1000\,MeV, is close 
to the radio galaxy 3C\,111, and plausibly associated with it; a new source, 
GRO\,J0426+3747, is also seen only above 
1000\,MeV, and has no obvious identification at other wavelengths.  Most of the 
flux $>$100\,MeV is from the third source, for which there is no evidence in the 
$>$1000\,MeV data.  The position obtained here for the third source is very near 
the catalog position of 3EG\,J0416+3650.  It  might be associated with the XRT 
source Swift\,J041554.3+364926, but the EGRET position uncertainty makes a 
firm association impossible.  The region under study is $8-10^\circ$ from the 
Galactic plane, so either 3EG\,J0416+3650 or GRO\,J0426+3747 (or both) could be 
local.

A recent stacking search found no evidence for $\gamma$--ray emmision from radio 
galaxies as a class \citep{Cil04}.  That paper specifically excluded 3C\,111 
from consideration 
because of its blazar-like superluminal motion, so there is no direct conflict 
with the results presented here.  That study made no systematic cut on the 
angle to the line-of-sight of the radio jet, so for most of the galaxies included, 
that angle was considerably larger than the $\sim$19$^\circ$ which has been 
estimated for 3C\,111 \citep{Kad08}.  Since a larger line-of-sight angle implies 
substantially smaller Doppler boosting and therefore lower $\gamma$--ray output, 
there is no obvious conflict between the present results and those of 
\citet{Cil04}.  A more significant challenge is posed by the lack of EGRET 
detection of $\gamma$--ray emission from the (much closer) AGN M87.  Superluminal 
motion has been detected in the jet knot HST-1, about 0.86\,arcsec downstream
from the radio core, both in the optical and radio bands
\citep{Bir99,Che07}.
Most estimates of our line-of-sight angle to its 
jet are in the range $30-45^\circ$.  If we assume that M87 is similar to 
3C\,111, it is not clear whether its Doppler 
boosting is sufficiently low to compensate for its much smaller distance, 
and thereby account for its non-detection by EGRET.

Our re-analysis of the available EGRET data supports the previous association of the source 
3EG\,J0416+3650 with the broad line radio galaxy 3C\,111, but with 3C\,111 
responsible for only a portion of the $\gamma$--ray emission.  Furthermore, it 
explains the relatively large positional offset noted in \citet{Ha99}.
We have compiled an historical SED of 3C\,111 which shows that the X--ray data 
may well extrapolate into the EGRET range, particularly during flares, which is 
in agreement with the intermittent nature of detections in the individual EGRET 
viewing periods. It is interesting to note that we detect 3C\,111 at almost 
exactly the $\gamma$--ray flux that is predicted by equation (12) in \citet{Ghi05}: 
$(1.41-14.1) \times 10^{-11}$\,erg\,s$^{-1}$\,cm$^{-2}$, scaling from the 
5\,GHz values in Table~\ref{tab:3c111_sed}.  Note that 3C\,111 (z=0.0485) would 
be the most distant radio galaxy detected in $\gamma$ rays, about twice as far 
as NGC\,6251 (Mukherjee et al. 2002; z=0.0247).

The detection of $\gamma$--ray emission from 3C\,111 further supports the hypothesis 
that radio galaxies may represent an important class of LAT sources.  GLAST 
was launched on 11 June, 2008; its LAT detector will continuously scan the 
entire sky over a 3-hr interval.  With a factor of 30 greater sensitivity than 
EGRET, and with a factor of $\sim$3 better PSF above 1\,GeV, it 
will be more efficient than EGRET for detecting transients over the whole sky.  
In the case of 3C\,111, GRO\,J0426+3747, and 3EG\,J0416+3650, the smaller PSF 
and greater effective area of the LAT will clearly separate these three sources.

\acknowledgements
{\small
We thank Goro Sato, Francesco Verrecchia, Svetlana Jorstad, and Chris Shrader for 
their help and advice in compiling the SED data for this study.  Steve Drake and 
Rachel Osten have kindly provided insight regarding X--ray emission from stars.  
MK has been supported by the NASA Postdoctoral Program at the 
Goddard Space Flight Center, administered by Oak Ridge Associated Universities 
through a contract with NASA.  This research has made use of data obtained from 
the High Energy Astrophysics Science Archive Research Center (HEASARC), 
provided by NASA's Goddard Space Flight Center, and the NASA/IPAC Extragalactic 
Database (NED), operated by the Jet Propulsion Laboratory, California Institute 
of Technology, under contract with NASA.  We have also made use of data products 
from the Two Micron All Sky Survey, a joint project of the University of 
Massachusetts and the Infrared Processing and Analysis Center, California 
Institute of Technology, funded by NASA and NSF.  Submillimeter flux ranges for 
3C\,111 were obtained from the on-line archives of the Submillimeter Array (SMA), 
a joint project between the Smithsonian Astrophysical Observatory and the 
Academia Sinica Institute of Astronomy and Astrophysics, funded by the Smithsonian
Institution and the Academia Sinica.
}

%\newpage

%\clearpage
\begin{deluxetable}{ccccccccccccccccc}
\tabletypesize{\scriptsize}
%\rotate
\setlength{\tabcolsep}{0.0035\textwidth} 
\tablecaption{EGRET detections of 3C\,111 and 3EG\,J0416+3650 discussed in the main text.  
         \label{tab:decomposition}}
\tablewidth{\textwidth}
\tablehead{
             &      &         &                              & \multicolumn{6}{c}{\hrulefill \, $>$100\,MeV \hrulefill} & \multicolumn{6}{c}{\hrulefill \, $>$1000\,MeV \hrulefill}\\
             &       &         &                              & \multicolumn{3}{c}{\hrulefill \, 3C\,111 \hrulefill} &  \multicolumn{3}{c}{\hrulefill \, J\,J0416+3650 \hrulefill} & \multicolumn{3}{c}{\hrulefill \, 3C\,111 \hrulefill} &  \multicolumn{3}{c}{\hrulefill \, J\,J0416+3650 \hrulefill}  \\
\colhead{VP$^{\rm a}$} & \colhead{Start Date} & \colhead{Days} & \colhead{Off-Axis$^{\rm b}$} & \colhead{TS$^{\rm c}$} & \colhead{Offset$^{\rm d}$} & \colhead{Flux$^{\rm e}$}  & \colhead{TS$^{\rm f}$} & \colhead{Offset$^{\rm d}$} & \colhead{Flux$^{\rm e}$}  & \colhead{TS$^{\rm c}$} & \colhead{Offset$^{\rm d}$} & \colhead{Flux$^{\rm e}$}  & \colhead{TS$^{\rm f}$} & \colhead{Offset$^{\rm d}$} & \colhead{Flux$^{\rm e}$}  \\
             &             &    & \colhead{[deg]}              &              & \colhead{[deg]}      &                           &              & \colhead{[deg]}      &                           &              & \colhead{[deg]}      &                           &              & \colhead{[deg]} 
}
\startdata
0002&  04-22-1991 &6&    24.82   &       &   -    & $<$33 &  18.3 & 1.2   &  60    &       &   -  &  $<$5  &  0.9 &   -  & $<$11  \\
0150&  11-28-1991 &14&   10.01   &       &   -    & $<$11 &   3.8 &   -   & $<$19 & 10.8  & 0.43 &   2     &  0.8 &   -  & $<$3   \\
0310&  06-11-1992 &14&   20.79   &  10.2 &   0.2  &   26   &   0.6 &   -   & $<$25 &  8.9  & 0.19 &   3     &      &   -  & $<$3   \\
0390&  09-01-1992 &16&    5.45   &       &   -    & $<$13 &   1.6 &   -   & $<$23 &  1.3  &   -  &  $<$4  &      &   -  & $<$4   \\
3211&  02-08-1994 &7&    20.60   &       &   -    & $<$33 &  17.1 & 0.2   &  55    &       &   -  &  $<$8  &  2.4 &   -  & $<$11  \\
3215&  02-15-1994 &2&    20.60   &   9.0 &   0.5  &   64   &       &   -   & $<$48 &       &   -  &  $<$20 &      &   -  & $<$16  \\
3250&  04-26-1994 &14&   14.47   &   1.7 &   -    & $<$24 &       &   -   & $<$14 &  4.1  & 0.09 &   2     &  0.9 &   -  & $<$5   \\
4270&  08-22-1995 &16&    7.91   &   3.0 &   -    & $<$38 &   0.1 &   -   & $<$22 &  0.2  &   -  &  $<$3  &  3.8 &   -  & $<$4   \\
P1234&   --       &--&   ---     &   2.7 &   -    & $<$9  &  16.2 & 0.1   &  10    &  8.7  & 0.10 &   1     &  3.1 &   -  & $<$1   \\
9185&  04-25-2000 &14&   25.65   &   0.1 &   -    & $<$78 &       &   -   & $<$75 &  0.3  &   -  &  $<$20 &      &   -  & $<$8   \\
\enddata
%% Text for table notes should follow after the \enddata but before
%% the \end{deluxetable}. Make sure there is at least one \tablenotemark
%% in the table for each \tablenotetext.
%\tablecomments{ Comment here }
\tablenotetext{a}{Only VPs with source detections are shown in this table. Seven additional VPs  showed no indication of
emission from either 3C\,111 or 3EG\,J0416+3650 (see text). VP0390 (TS=0) is included here because it was the best exposure to the relevant sky region.}
\tablenotetext{b}{Off-axis angle of 3C\,111}
\tablenotetext{c}{Maximum test statistic near the position of 3C\,111}
\tablenotetext{d}{Angle between the original position and the position of maximum TS}
\tablenotetext{e}{Flux in units of $10^{-8}$\,ph\,cm$^{-2}$\,s$^{-1}$}
\tablenotetext{f}{Maximim test statistic near the position of 3EG\,J0416+3650}
\end{deluxetable}

\begin{deluxetable}{ccccc}
\tabletypesize{\small}
%\rotate
\tablewidth{0pt}
%\setlength{\tabcolsep}{0.025\textwidth}
%\centering
\tablecaption{$>$1000\,MeV excesses $\ge$2$\sigma$ in 17 maps (5$^\circ$ by 5$^\circ$) of the field around 3EG\,J0416+3650.  \label{tab:2-sigma_table}}
\tablehead{
\colhead{VP} & \colhead{Full} & \colhead{3C\,111} & \colhead{0426+3747} & \colhead{Background}\\
             & \colhead{Field$^{\rm a}$}& \colhead{Field$^{\rm b}$}   & \colhead{Field$^{\rm c}$}            & \colhead{Field$^{\rm d}$}     
}
\startdata
0002 &            0   &        0  &      0  &             0 \\
0005 &            s   &        0  &      s  &             0 \\
0010 &            1   &        0  &      0  &             1 \\
0150 &            2   &        1  &      1  &             0 \\
0310 &            1   &        1  &      0  &             0 \\
0360 &            1   &        0  &      0  &             1 \\
0365 &            0   &        0  &      0  &             0 \\
0390 &            1   &        0  &      0  &             1 \\
2130 &            1   &        0  &      0  &             1 \\
2210 &            1   &        0  &      1  &             1 \\
3211 &            2   &        0  &      1  &             1 \\
3215 &            s   &        0  &      s  &             0 \\
3250 &            4   &        1  &      1  &             2 \\
4120 &            2   &        0  &      0  &             2 \\
4260 &            0   &        0  &      0  &             0 \\
4270 &            1   &        0  &      0  &             1 \\
8290 &            1   &        0  &      1  &             0 \\
\hline
Sum		& $18 + 2s$         & $3$                & $5 + 2s$           & $11$\\
Rate	& 10\% & 18\% & 29\% & 7\% 
\enddata
%% Text for table notes should follow after the \enddata but before
%% the \end{deluxetable}. Make sure there is at least one \tablenotemark
%% in the table for each \tablenotetext.
\tablecomments{~s (in VPs 0005, 3215, columns 2 and 4) -  a single photon 
$>$1000\,MeV very near the source position, probably significant because of the short 
exposures - see text; $^{\rm a}$ 187 PSFs; $^{\rm b}$ 17 PSFs; $^{\rm c}$ 17 PSFs; $^{\rm d}$ 153 PSFs}
\end{deluxetable}

\begin{deluxetable}{clrccc}
\tabletypesize{\scriptsize}
%\rotate
\setlength{\tabcolsep}{0.016\textwidth} 
\tablecaption{3C\,111 SED data 
         \label{tab:3c111_sed}}
\tablewidth{1.0\textwidth}
\tablehead{
Frequency$^{\rm a}$ & Band & Instrument & Flux$^{\rm b}$ & Flux Error & Reference$^{\rm c}$\\
}
\startdata
$(0.24 - 24.1) \times 10^{23}$	&	$>100$\,MeV	&	EGRET(max)	&$1.92 \times 10^{-10}$	&$7.07 \times 10^{-11}$	&    (1)     \\
$(2.41 - 24.1) \times 10^{23}$	&	$>1000$\,MeV	&	EGRET(max)	&$1.11 \times 10^{-10}$	&$7.00 \times 10^{-11}$	&    (1)     \\
$(2.41 - 24.1) \times 10^{23}$	&	$>1000$\,MeV	&	EGRET(avge)	&$2.90 \times 10^{-11}$	&$1.50 \times 10^{-11}$	&    (1)     \\
$(0.24 - 24.1) \times 10^{23}$	&	$>100$\,MeV	&	EGRET(avge)	&$3.00 \times 10^{-11}$	&$1.90 \times 10^{-11}$	&    (1)     \\
$(0.24 - 24.1) \times 10^{23}$	&	$>100$\,MeV	&	EGRET(VP3215)	&$4.28 \times 10^{-10}$	&$2.06 \times 10^{-10}$	&    (1)     \\
$(2.41 - 24.1) \times 10^{23}$	&	$>1000$\,MeV	&	EGRET(VP3215)	&$<1.53 \times 10^{-10}$ &			&    (1)     \\
$(2.41 - 7.23) \times 10^{20}$	&	$1-3$\,MeV	&	COMPTEL		&$<1.30 \times 10^{-10}$ &			& (2) \\
$(7.23 - 24.1) \times 10^{20}$	&	$3-10$\,MeV	&	COMPTEL		&$<1.30 \times 10^{-10}$ &			&    (2)     \\
$(2.41 - 7.23) \times 10^{21}$	&	$10-30$\,MeV	&	COMPTEL		&$<3.00 \times 10^{-10}$ &			&    (2)     \\
$(1.81 - 2.41) \times 10^{20}$	&	$0.75-1$\,MeV	&	COMPTEL		&$<2.90 \times 10^{-10}$ &			&    (2)     \\
$(1.20 - 3.60) \times 10^{19}$	&	$50-150$\,keV	&	OSSE		&$3.52 \times 10^{-11}$	&$6.56 \times 10^{-12}$	& (3) \\
$(9.70 - 24.0) \times 10^{18}$	&	$40-100$\,keV	&	ISGRI		&$8.10 \times 10^{-11}$	&$7.40 \times 10^{-12}$	& (4) \\
$(4.80 - 9.70) \times 10^{18}$	&	$20-40$\,keV	&	ISGRI		&$6.27 \times 10^{-11}$	&$5.70 \times 10^{-12}$	&    (4)      \\
$(3.39 - 47.2) \times 10^{18}$	&	$14-195$\,keV	&	BAT		&$1.25 \times 10^{-10}$	&$9.00 \times 10^{-12}$	& (5) \\
$(4.80 - 19.3) \times 10^{18}$	&	$2-80$\,keV	&	BeppoSax	&$4.99 \times 10^{-11}$	&$9.98 \times 10^{-12}$	&  (6)  \\
$(4.80 - 24.0) \times 10^{17}$	&	$2-10$\,keV	&	BeppoSax	&$2.49 \times 10^{-11}$	&$4.98 \times 10^{-12}$	&   (6)       \\
$(1.20 - 4.80) \times 10^{17}$	&	$0.5-2$\,keV	&	BeppoSax	&$4.36 \times 10^{-12}$	&$8.72 \times 10^{-13}$	&   (6)      \\
$(4.80 - 24.0) \times 10^{17}$	&	$2-10$\,keV	&	XMM		&$6.57 \times 10^{-11}$	&$3.94 \times 10^{-13}$	&   (6)      \\
$(1.20 - 4.80) \times 10^{17}$	&	$0.5-2$\,keV	&	XMM		&$1.38 \times 10^{-11}$	&$3.94 \times 10^{-13}$	&   (6)      \\
$(4.80 - 24.0) \times 10^{17}$	&	$2-10$\,keV	&	ASCA		&$3.92 \times 10^{-11}$	&$7.84 \times 10^{-12}$	&   (6)      \\
$(1.20 - 4.80) \times 10^{17}$	&	$0.5-2$\,keV	&	ASCA		&$7.42 \times 10^{-12}$	&$1.48 \times 10^{-12}$	&   (6)      \\
$1.14 \times 10^{15}$		&	UVW1		&	UVOT		&$<6.98 \times 10^{-13}$&	& (7)\\
$8.57 \times 10^{14}$		&	U		&	UVOT		&$<6.95 \times 10^{-13}$&	&    (7)     \\
$6.92 \times 10^{14}$		&	B		&	UVOT		&$1.00 \times 10^{-12}$	&$(+1.0,-0.5) \times 10^{-12}$	&    (7)     \\
$5.55 \times 10^{14}$		&	V		&	UVOT		&$2.30 \times 10^{-12}$	&$(+1.1,-0.7) \times 10^{-13}$	&    (7)     \\
$4.29 \times 10^{14}$		&	R		&	Liverpool	&$(4.22 - 7.25) \times 10^{-12}$	&	& (8) \\
$2.40 \times 10^{14}$		&	J		&	2MASS		&$1.19 \times 10^{-11}$	&$4.00 \times 10^{-13}$	& (9), (10) \\
$1.82 \times 10^{14}$		&	H		&	2MASS		&$1.59 \times 10^{-11}$	&$4.62 \times 10^{-13}$	&     (9), (10)      \\
$1.36 \times 10^{14}$		&	K		&	2MASS		&$1.92 \times 10^{-11}$	&$3.66 \times 10^{-13}$	&     (9), (10)      \\
$2.50 \times 10^{13}$		&	$12$\,micron	&	IRAS		&$2.43 \times 10^{-11}$	&$3.65 \times 10^{-12}$	& (11) \\
$1.20 \times 10^{13}$		&	$25$\,micron	&	IRAS		&$2.69 \times 10^{-11}$	&$4.04 \times 10^{-12}$	&     (11)      \\
$5.00 \times 10^{12}$		&	$60$\,micron	&	IRAS		&$1.61 \times 10^{-11}$	&$2.42 \times 10^{-12}$	&     (11)      \\
$3.40 \times 10^{11}$		&	$0.85$\,mm	&	SMA		&$(5.27 - 28.4) \times 10^{-12}$ & & (12)		\\
$2.30 \times 10^{11}$		&	$1.3$\,mm	&	SMA		&$(0.94 - 25.9) \times 10^{-12}$ & &    (12)      \\
$9.00 \times 10^{10}$		&	$3$\,mm		&	SMA		&$(0.91 - 3.66) \times 10^{-12}$ & &    (12)      \\
$1.50 \times 10^{10}$		&	$2$\,cm		&	UMRAO		&$(0.30 - 1.05) \times 10^{-12}$ & & (13) \\
$8.00 \times 10^{09}$		&	$3.6$\,cm	&	UMRAO		&$(3.20 - 6.40) \times 10^{-13}$ & &    (13)       \\
$5.00 \times 10^{09}$		&	$6$\,cm		&	UMRAO		&$(3.00 - 4.50) \times 10^{-13}$ & &     (13)      \\
\enddata
%\tablecomments{ Comment here }
\tablenotetext{a}{Frequency in units of Hz}
\tablenotetext{b}{Flux in units of erg\,cm$^{-2}$\,s$^{-1}$}
\tablenotetext{c}{References: (1) this work; (2) \cite{Mai97}; (3) \cite{Jo87}; (4) \cite{Be06}; (5) \cite{Tu07}; (6) \cite{Kad05}; 
(7) F. Verrecchia, priv. comm.; (8) S. Jorstad, priv. comm.; (9) 2MASS archive; (10) \cite{Sr06}; (11) \cite{Go88}; 
(12) SMA archive ({\tt http://cfa-www.harvard.edu/rtdc/index-sma.html}); (13) \cite{Kad08}
}
\end{deluxetable}

\begin{deluxetable}{clrccc}
\tabletypesize{\scriptsize}
%\rotate
\setlength{\tabcolsep}{0.0242\textwidth} 
\tablecaption{3EG\,J0416+3650 SED data 
         \label{tab:J0416_sed}}
\tablewidth{1.0\textwidth}
\tablehead{
Frequency$^{\rm a}$ & Band & Instrument & Flux$^{\rm b}$ & Flux Error & Reference$^{\rm c}$\\
}
\startdata
$(0.24 - 24.1) \times 10^{23}$	&	$>100$\,MeV	&	EGRET(avge)	&$7.03 \times 10^{-11}$ &$1.89 \times 10^{-11}$	& (1) \\
$(0.24 - 24.1) \times 10^{24}$	&	$>100$\,MeV	&	EGRET(VP3211)	&$4.42 \times 10^{-10}$	&$1.29 \times 10^{-10}$ &    (1)     \\
$(3.39 - 47.2) \times 10^{18}$	&	$14-195$\,keV	&	BAT		&$<1.78 \times 10^{-11}$&			& (2) \\
$(4.82 - 24.1) \times 10^{17}$	&	$2-10$\,keV	&	XRT		&$1.05 \times 10^{-13}$	&$5.24 \times 10^{-14}$	& (1) \\
$1.48 \times 10^{15}$		&	UVW2		&	UVOT(observed)	&$4.43 \times 10^{-14}$	&$1.48 \times 10^{-14}$	&   (1)      \\
$1.34 \times 10^{15}$		&	UVM2		&	UVOT(observed)	&$5.38 \times 10^{-14}$	&$1.35 \times 10^{-14}$	&   (1)      \\
$1.14 \times 10^{15}$		&	UVW1		&	UVOT(observed)	&$1.94 \times 10^{-13}$	&$2.28 \times 10^{-14}$	&   (1)      \\
$8.57 \times 10^{14}$		&	U		&	UVOT(observed)	&$7.80 \times 10^{-13}$	&$3.43 \times 10^{-14}$	&   (1)      \\
$6.93 \times 10^{14}$		&	B		&	UVOT(observed)	&$9.29 \times 10^{-13}$	&$5.57 \times 10^{-14}$	&   (1)      \\
$5.55 \times 10^{14}$		&	V		&	UVOT(observed)	&$1.13 \times 10^{-12}$	&$9.40 \times 10^{-14}$	&   (1)      \\
$1.48 \times 10^{15}$		&	UVW2		&	UVOT(dereddened)&$5.16 \times 10^{-12}$	&$1.72 \times 10^{-12}$	&   (1)      \\
$1.34 \times 10^{15}$		&	UVM2		&	UVOT(dereddened)&$8.06 \times 10^{-12}$	&$2.02 \times 10^{-12}$	&   (1)      \\
$1.14 \times 10^{15}$		&	UVW1		&	UVOT(dereddened)&$6.25 \times 10^{-12}$	&$7.38 \times 10^{-13}$	&   (1)      \\
$8.57 \times 10^{14}$		&	U		&	UVOT(dereddened)&$1.06 \times 10^{-11}$	&$4.67 \times 10^{-13}$	&   (1)      \\
$6.93 \times 10^{14}$		&	B		&	UVOT(dereddened)&$8.15 \times 10^{-12}$	&$4.89 \times 10^{-13}$	&   (1)      \\
$5.55 \times 10^{14}$		&	V		&	UVOT(dereddened)&$6.22 \times 10^{-12}$	&$5.17 \times 10^{-13}$	&   (1)      \\
$4.29 \times 10^{14}$		&	R		&	2MASS		&$<4.58 \times 10^{-12}$&			& (3),(4) \\
$2.40 \times 10^{14}$		&	J		&	2MASS		&$2.66 \times 10^{-12}$	&$2.66 \times 10^{-13}$	&    (3),(4)     \\
$1.82 \times 10^{14}$		&	H		&	2MASS		&$1.73 \times 10^{-12}$	&$2.60 \times 10^{-13}$	&    (3),(4)     \\
$1.36 \times 10^{14}$		&	K		&	2MASS		&$1.57 \times 10^{-12}$	&$1.57 \times 10^{-13}$	&    (3),(4)     \\
$1.40 \times 10^{09}$		&	$20$\,cm	&	NVSS		&$<1.40 \times 10^{-17}$&			& (5) \\
\enddata
%\tablecomments{ Comment here }
\tablenotetext{a}{Frequency in units of Hz}
\tablenotetext{b}{Flux in units of erg\,cm$^{-2}$\,s$^{-1}$}
\tablenotetext{c}{References: (1) this work; (2) \cite{Tu07}; (3) 2MASS archive; (4) \cite{Sr06}; (5) \cite{Con98}}
\end{deluxetable}

\end{document}